\documentclass{PoS}

\title{Observation of $B_s^{0} \to D_s^{*-} \pi^{+}$, $B_s^{0} \to D_s^{(*)-} \rho^{+} $ and $B_s^{0} \to D_s^{(*)+} D_s^{(*)-}$ and Estimate of $\Delta \Gamma_{CP}$ at Belle}

\ShortTitle{Observation of $B_s^{0} \to D_s^{*-} \pi^{+}$, $B_s^{0} \to D_s^{(*)-} \rho^{+} $ and $B_s^{0} \to D_s^{(*)+} D_s^{(*)-}$ and Estimate of $\Delta \Gamma_{CP}$}

\author{\speaker{Sevda Esen}%
        \thanks{For Belle Collaboration.}\\
       University of Cincinnati\\
       E-mail: \email{esens@mail.uc.edu}}

\abstract{The large data sample being recorded with the Belle detector at the $\Upsilon$(5S) energy provides a unique opportunity to study the less-well-known $B_s^{0}$ meson decays. Following our recent measurement of $B_s^{0}\to D_s^{-}\pi^{+}$ in a sample of 23.6~fb$^{-1}$, we extend the analysis to include decays with photons in the final state. Using the same sample, we report the first observation of three other dominant exclusive $B_s^{0}$ decays, in the modes $B_s^{0}\to D_s^{*-}\pi^+$,  $B_s^{0}\to D_s^{-}\rho^+$  and $B_s^{0}\to D_s^{*-}\rho^+$.  We measure their respective branching fractions and, using helicity-angle distributions, the longitudinal polarization fraction of the $B_s^{0}\to D_s^{*-}\rho^+$ decay.

We also present a measurement of the branching fractions for the decays $B_s^{0} \to D_s^{(*)+}D_s^{(*)-}$. In the heavy quark limit, this branching fraction is directly related to the width difference between the $B_s$ $CP$-even and $CP$-odd eigenstates.}

\FullConference{35th International Conference of High Energy Physics\\
		 July 22-28, 2010\\
		 Paris, France}

\begin{document}

\section{Introduction}
Beginning in 2005, the Belle experiment running KEKB  $e^{+}e^{-}$ collider \cite{Belle-KEKB} has recorded several data sets at the center-of-mass energy corresponding to the  $\Upsilon(5S)$ resonance. Belle has used this data sets to measure several $B_{s}^{0}$ properties and branching fractions. A total of 120 fb$^{-1}$ at the $\Upsilon(5S)$  ($\sqrt{s} \approx $10.87 GeV) has been recorded. The results presented here correspond to the first 23.6 fb$^{-1}$. 

The total $e^{+}e^{-}\to b\bar{b}$ cross section at the $\Upsilon(5S)$ energy was measured to be $\sigma_{b\bar{b}} = (302 \pm 14)$ pb \cite{AD98, CLEO75}, with the fraction $f_{s}= \sigma(e^{+}e^{-} \to B_{s}^{(*)}\bar{B}_{s}^{(*)} )/\sigma_{b\bar{b}} = (19.3 \pm 2.9) \% $ \cite{PDG}. The dominant $B_{s}^{0}$ production mode is  $e^{+}e^{-} \to B_{s}^{*}\bar{B}_{s}^{*}$, with a fraction $f_{B_{s}^{*}\bar{B}_{s}^{*}} = (90.1^{+3.8} _{-4.0}\pm 0.2 )\%$  of the  $b\bar{b} \to B_{s}^{(*)}\bar{B}_{s}^{(*)}$ events \cite{RL102}. Thus for 23.6 fb$^{-1}$ the total number of $e^{+}e^{-} \to B_{s}^{*}\bar{B}_{s}^{*}$ events  is $(1.24\pm0.2)\times 10^{6}$.

All signal $B_{s}^{0}$ decays are fully reconstructed from final-state particles using two quantities: the beam-energy-constrained mass  $M^{}_{bc}=\sqrt{E^2_b - p^2_B}$, and the energy difference $\Delta E= E^{}_B-E^{}_b$, where $p_B$ and $E_B$ are the reconstructed momentum and energy of the $B_{s}^{0}$ candidate, and $E_b$ is the beam energy. These quantities are evaluated in the $e^{+}e^{-}$ center-of-mass frame. Although the $B_{s}^{*}$ always decays to $B_{s}^{0}\gamma$, the $\gamma$ is not reconstructed because of its extremely low momentum. 

\section{Observation of $B_{s}^{0} \to D_{s}^{*-}\pi^{+}$ and $D_{s}^{(*)-}\rho^{+}$ Decays and Polarization Measurement of $B_{s}^{0} \to D_{s}^{*-}\rho^{+}$ }

Three CKM-favored decays with relatively large branching fractions, $B_{s}^{0} \to D_{s}^{*-}\pi^{+}$ and $D_{s}^{(*)-}\rho^{+}$, have been observed recently by Belle \cite{RL104}.  Three $D_{s}^{+} $ decay modes  are considered: $\phi( \to K^{+}K^{-})\pi^{+}$, $K_{S}(\to \pi^{+}\pi^{-})K^{+}$ and $K^{*0}( \to K^{+}\pi^{-})K^{+}$. Since only four charged tracks and up to one $\gamma$ and $\pi^{0}$ are required, these final states have relatively large signals. The continuum events are removed using the ratio of the second to zeroth Fox-Wolfram moments \cite{R2}. This ratio differs for spherical $B$ events and jet-like continuum events.

\begin{figure}[htb]
\includegraphics[trim= 0.0in 0.0in 0.0in 0.0in, clip, width=0.52\textwidth]{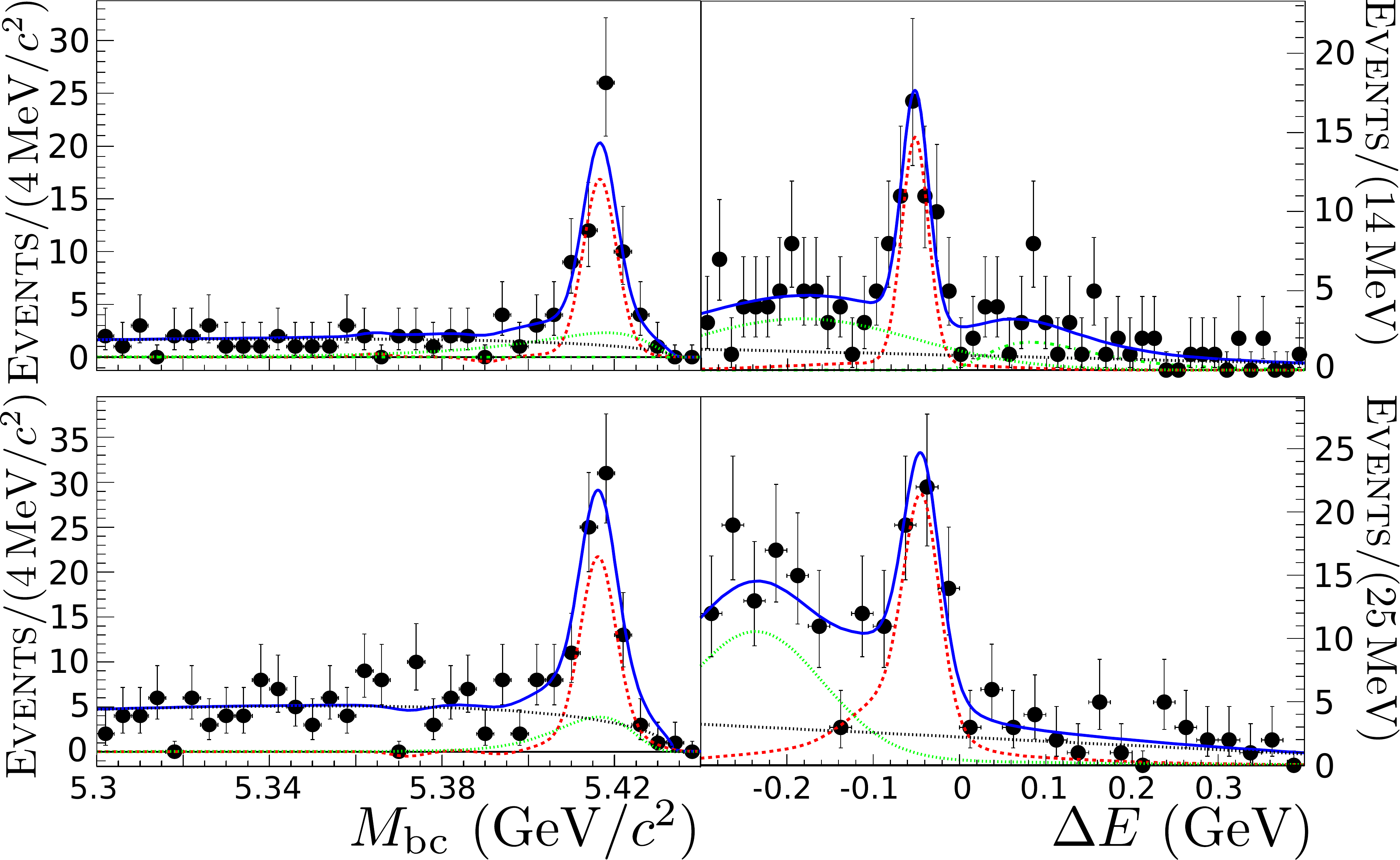}
\includegraphics[trim= 0.0in 0.0in 0.0in 0.0in, clip, width=0.48\textwidth]{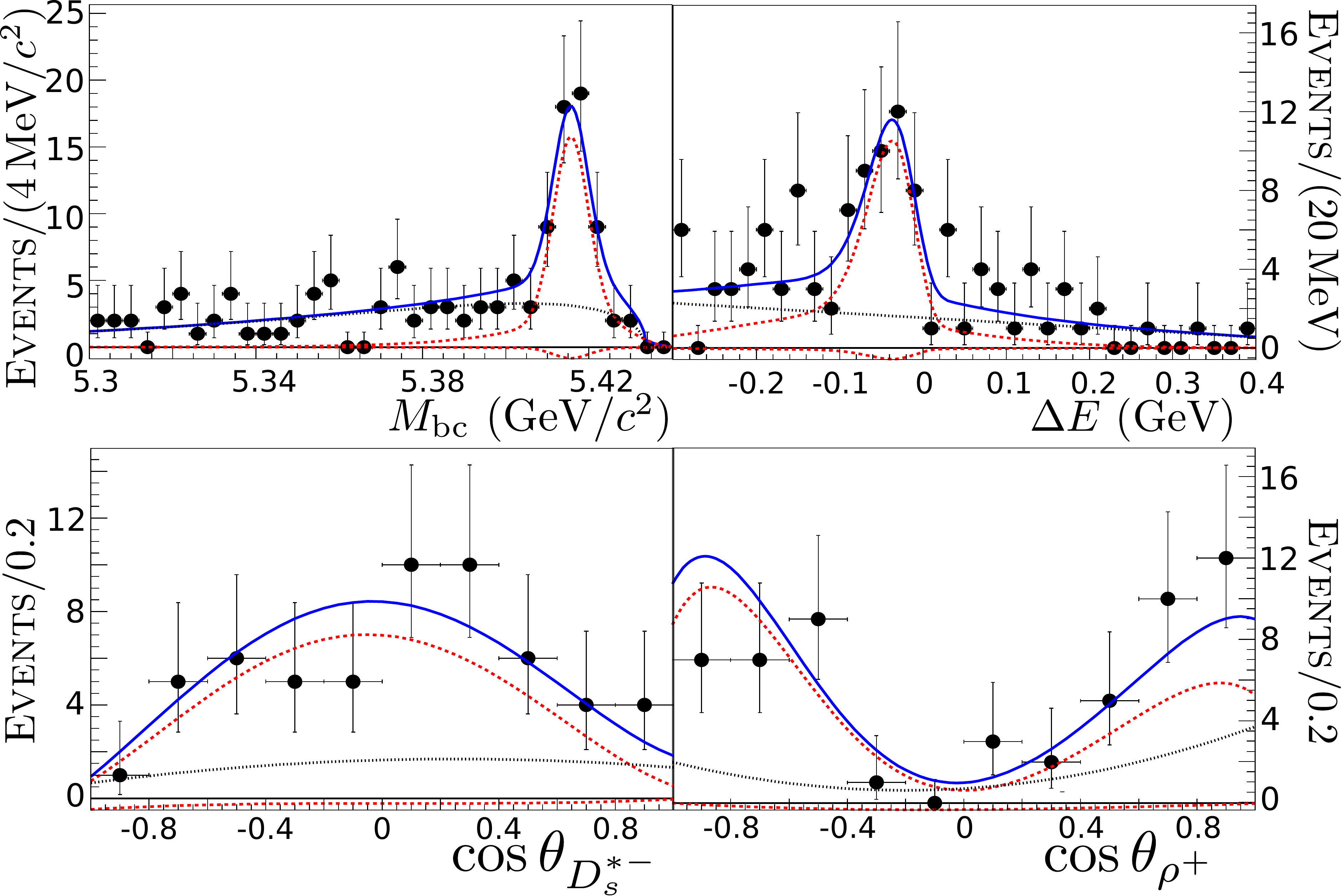}
\caption{Projections of $B_{s}^{*}\bar{B}_{s}^{*}$ signal region in $M_{bc}$ and $\Delta E$  for fits of $B_{s}^{0}$ to  $D_{s}^{*-}\pi^{+}$(top-left), $D_{s}^{-}\rho^{+}$ (bottom-left), and  $D_{s}^{*-}\rho^{+}$ (top-right).  The bottom-right  figure shows the helicity distributions for $D_{s}^{*-}\rho^{+}$ mode. The solid-blue line represents the total fit, while the red-dashed(black-dotted) curve is the signal(background). }
\label{fig:dsh}
\end{figure}

Only one $B_{s}^{0}$ candidate is allowed per event. This candidate is chosen based on  the intermediate-particle reconstructed masses. The $M_{bc}$ and $\Delta E$ distributions of the selected $B_{s}^{0}$ candidates are shown in Figure \ref{fig:dsh}. For the $B_{s}^{0}\to D_{s}^{*-}\rho^{+}$ candidates,  the helicity angles $\theta_{D_{s}^{*-}}$ and $\theta_{\rho^{-}}$ are also reconstructed. These are defined  as the angle between the $D_{s}^{-}$ or $\pi^{+}$ and the opposite direction of the $B_{s}^{0}$ in the $D_{s}^{*-}$ or $\rho^{-}$ rest frame. The distributions of $\cos\theta_{D_{s}^{*-}}$ and $\cos\theta_{\rho^{-}}$ are fitted to determine the longitudinal polarization fraction $f_{L}$ (see Table \ref{tab:results}).

\section{ Observation of $B_s \to D_s^{(*)-} D_{s}^{(*)+} $ Decays and a Determination of the $\Delta\Gamma_{s}$ }

Decays of $B_s \to  D_s^{(*)-} D_{s}^{(*)+} $ are interesting due to their large CP-even fraction.  The pure CP-even $D_{s}^{-}D_{s}^{+}$ state and predominantly CP-even $D_{s}^{*}D_{s}^{(*)}$ states are Cabibbo-favored and expected to dominate the width difference of the $B_{s}^{0}-\bar{B}_{s}^{0}$ system. In the heavy quark limit, assuming negligible CP violation, the relative width difference is 
${\Delta \Gamma_s^{CP}}/{\Gamma_s}=2 \mathcal{B} / (1- \mathcal{B})$,  where $\mathcal{B}$ is the total branching fraction of  $B_s \to  D_s^{(*)-} D_{s}^{(*)+} $ decays \cite{RA316}. 

\begin{figure}[htb]
\begin{center}
\includegraphics[trim= 0.0in 0.1in 0.0in 0.6in, clip, width=0.275\textwidth]{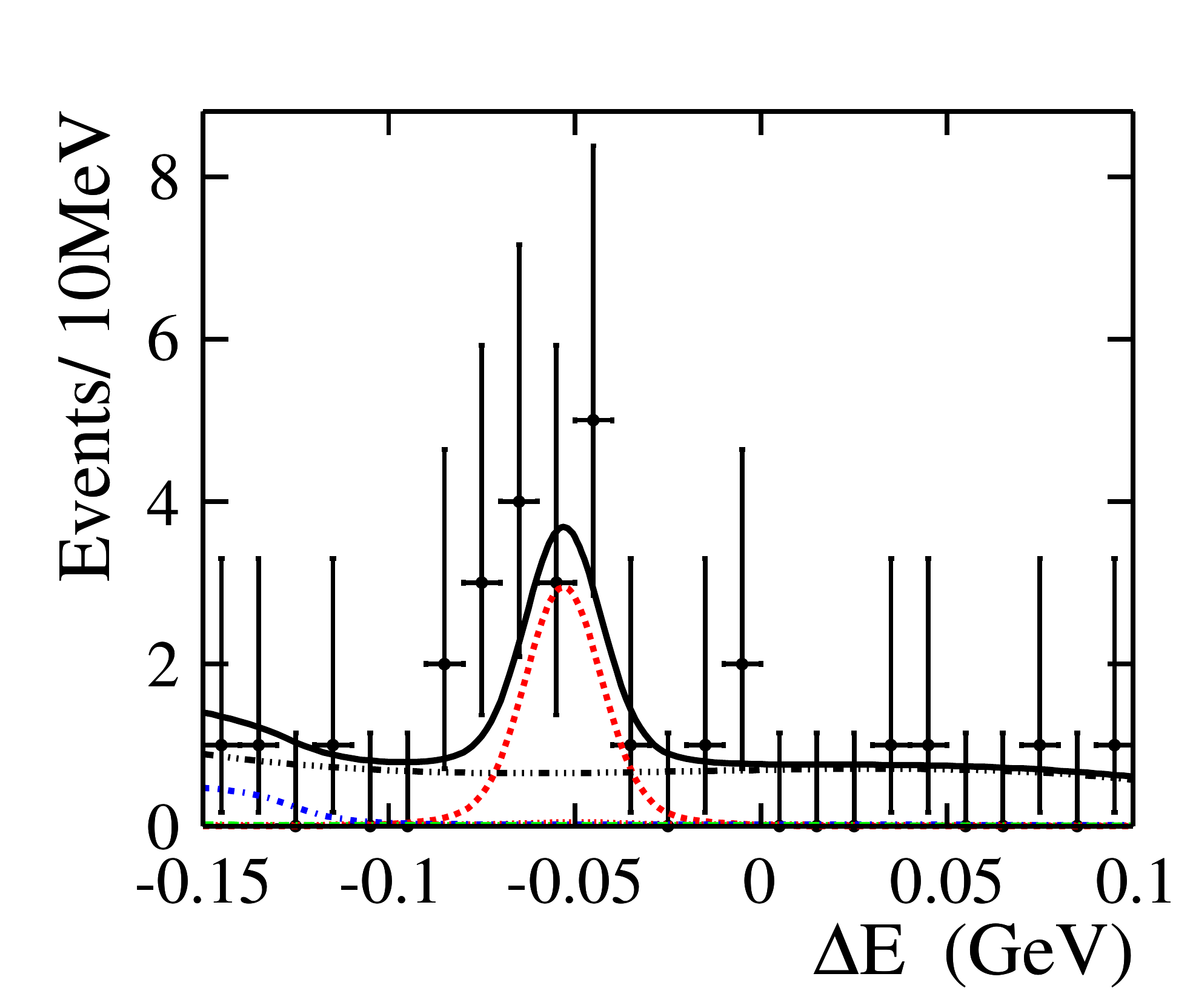}
\includegraphics[trim= 0.8in 0.1in 0.0in 0.67in, clip, width=0.245\textwidth]{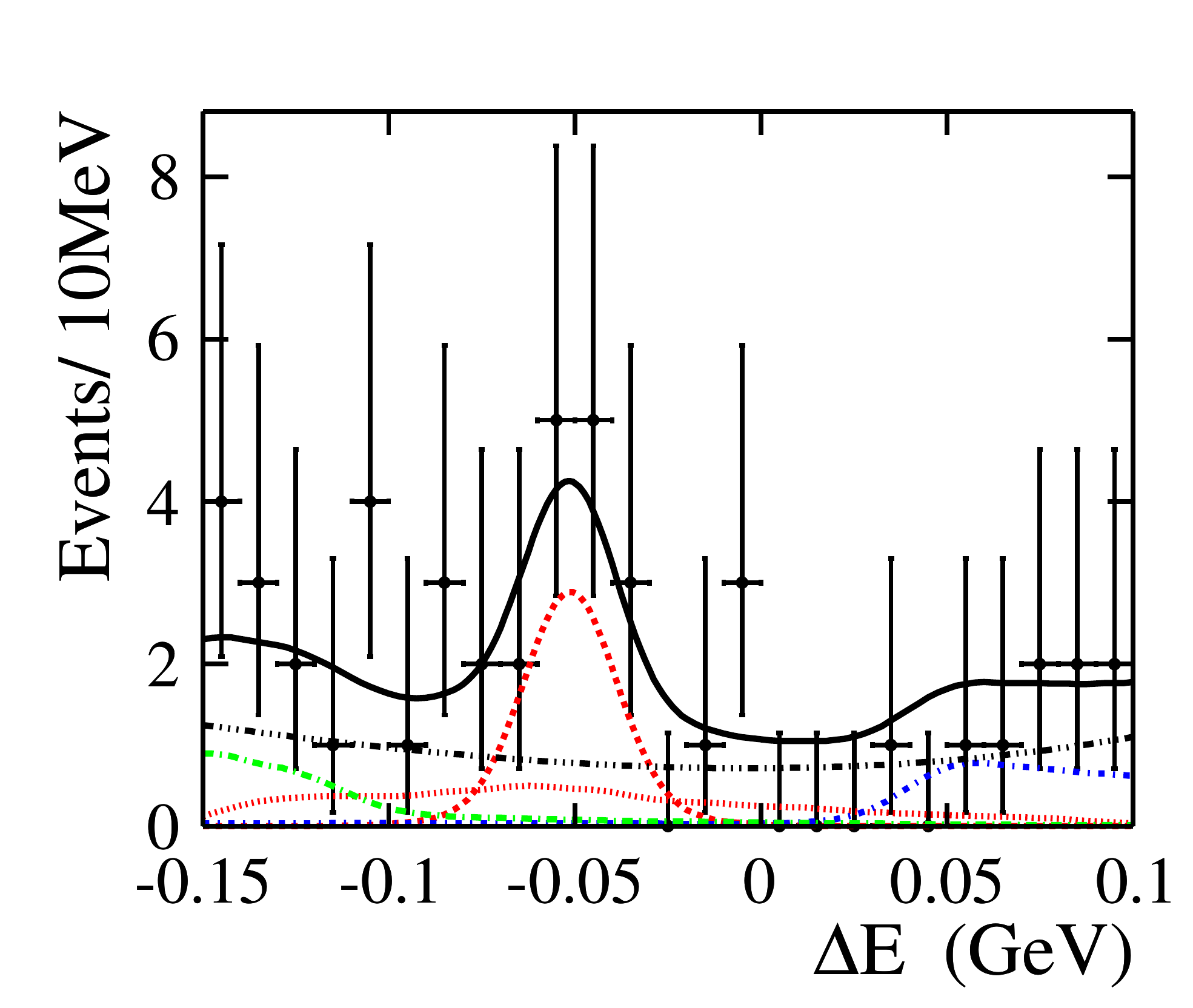}
\includegraphics[trim= 0.8in 0.1in 0.0in 0.6in, clip, width=0.245\textwidth]{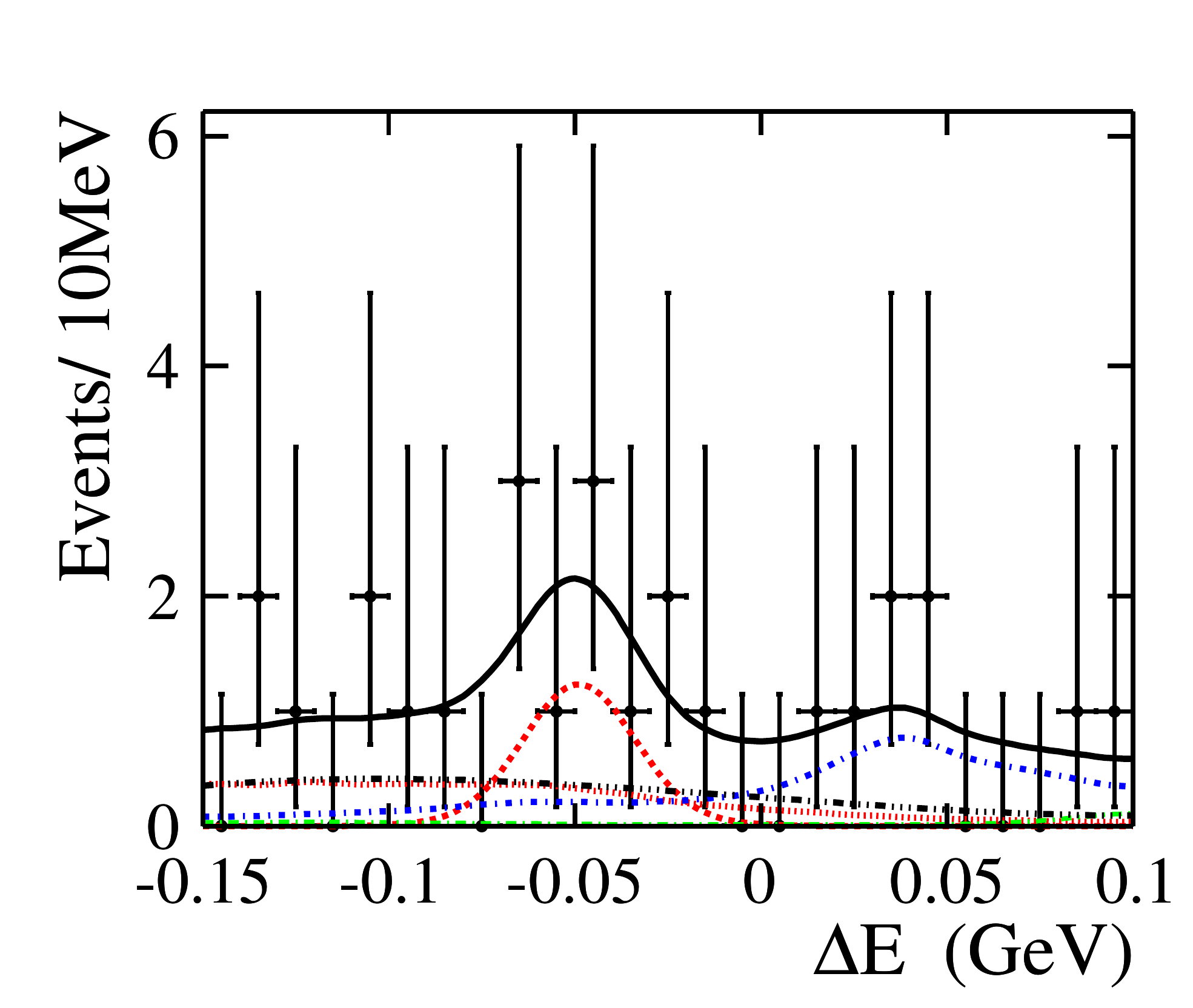}
\includegraphics[trim= 0.0in 0.0in 0.0in 0.7in, clip, width=0.275\textwidth]{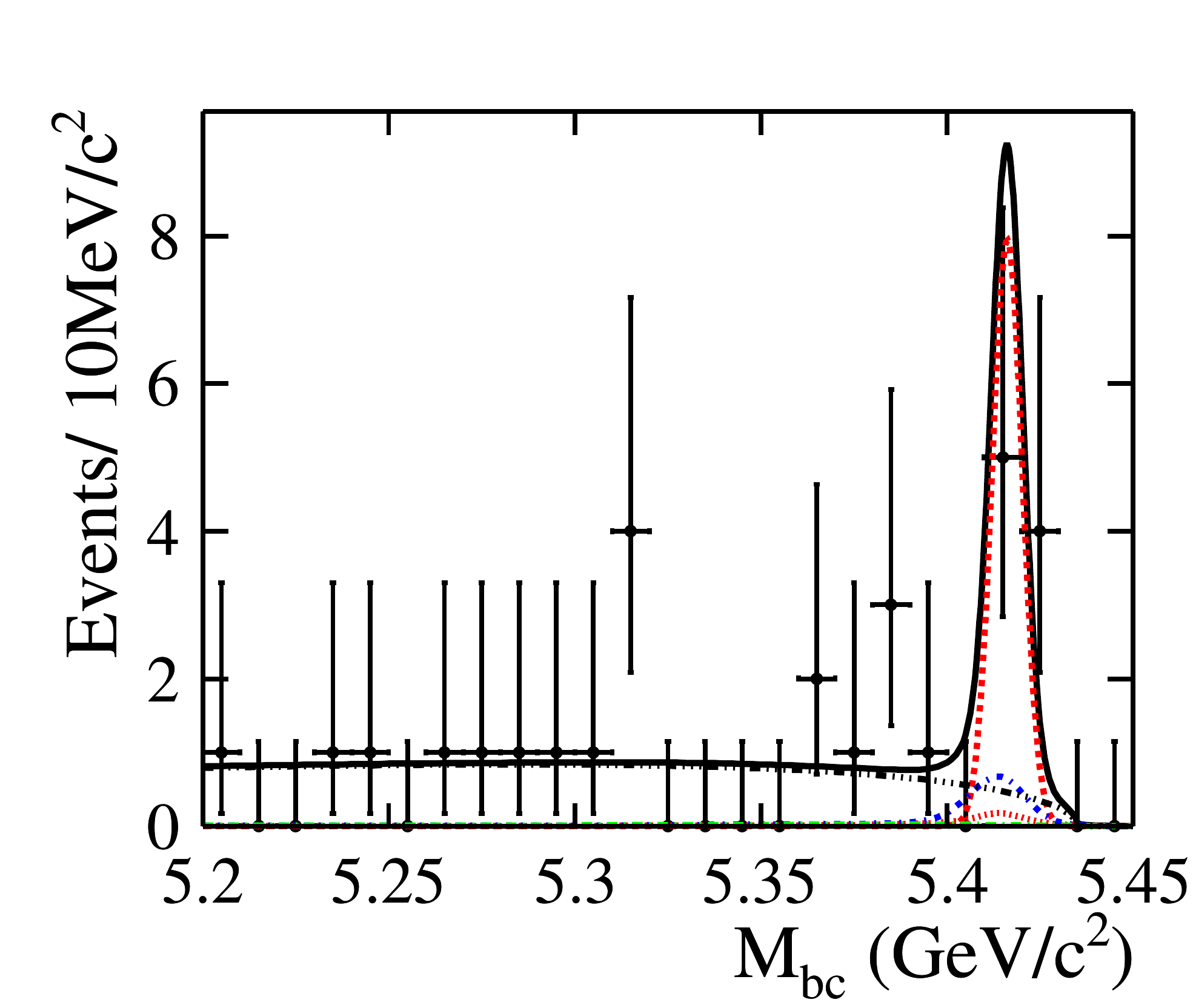}
\includegraphics[trim= 0.8in 0.0in 0.0in 0.7in, clip, width=0.245\textwidth]{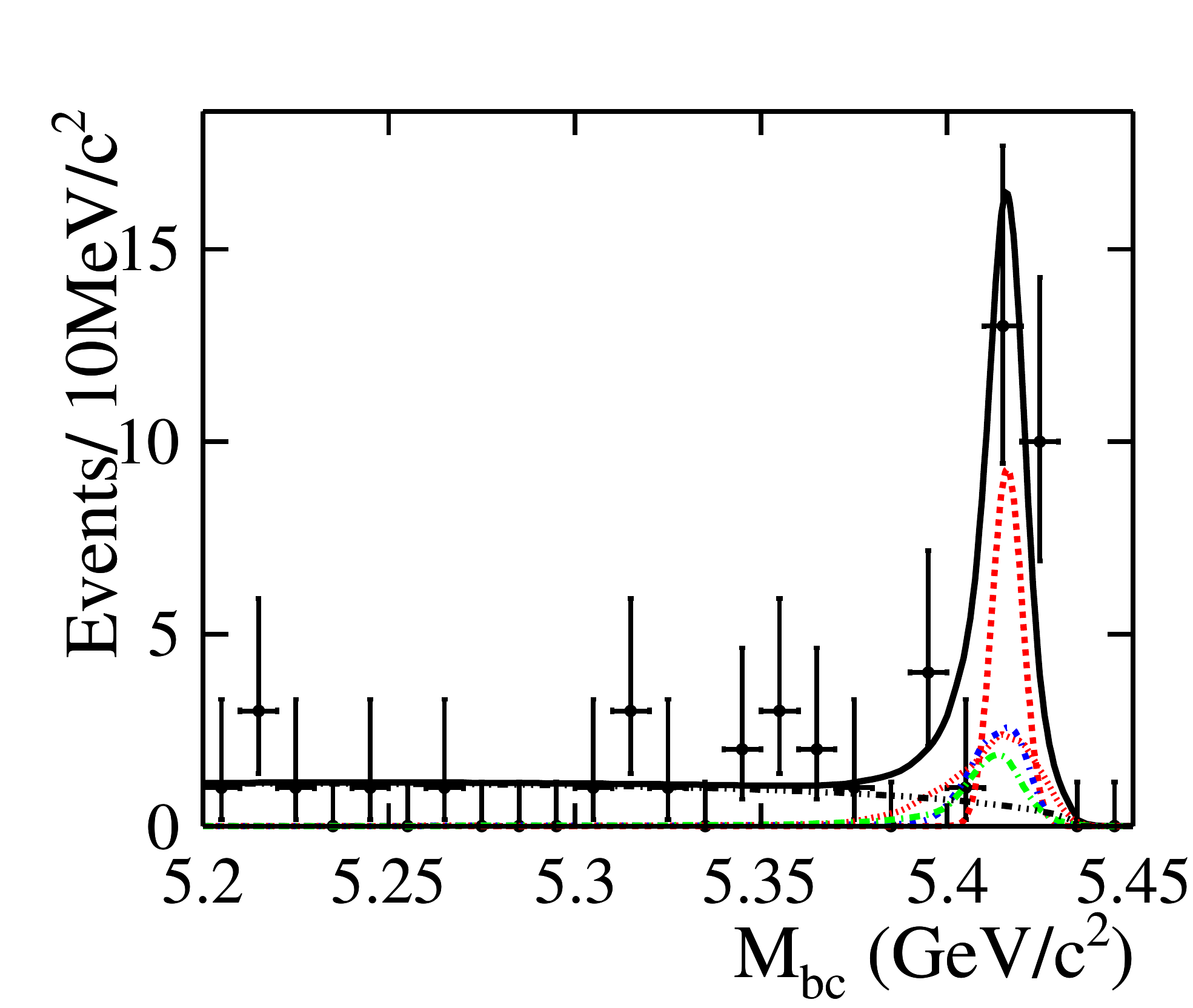}
\includegraphics[trim= 0.8in 0.0in 0.0in 0.7in, clip, width=0.245\textwidth]{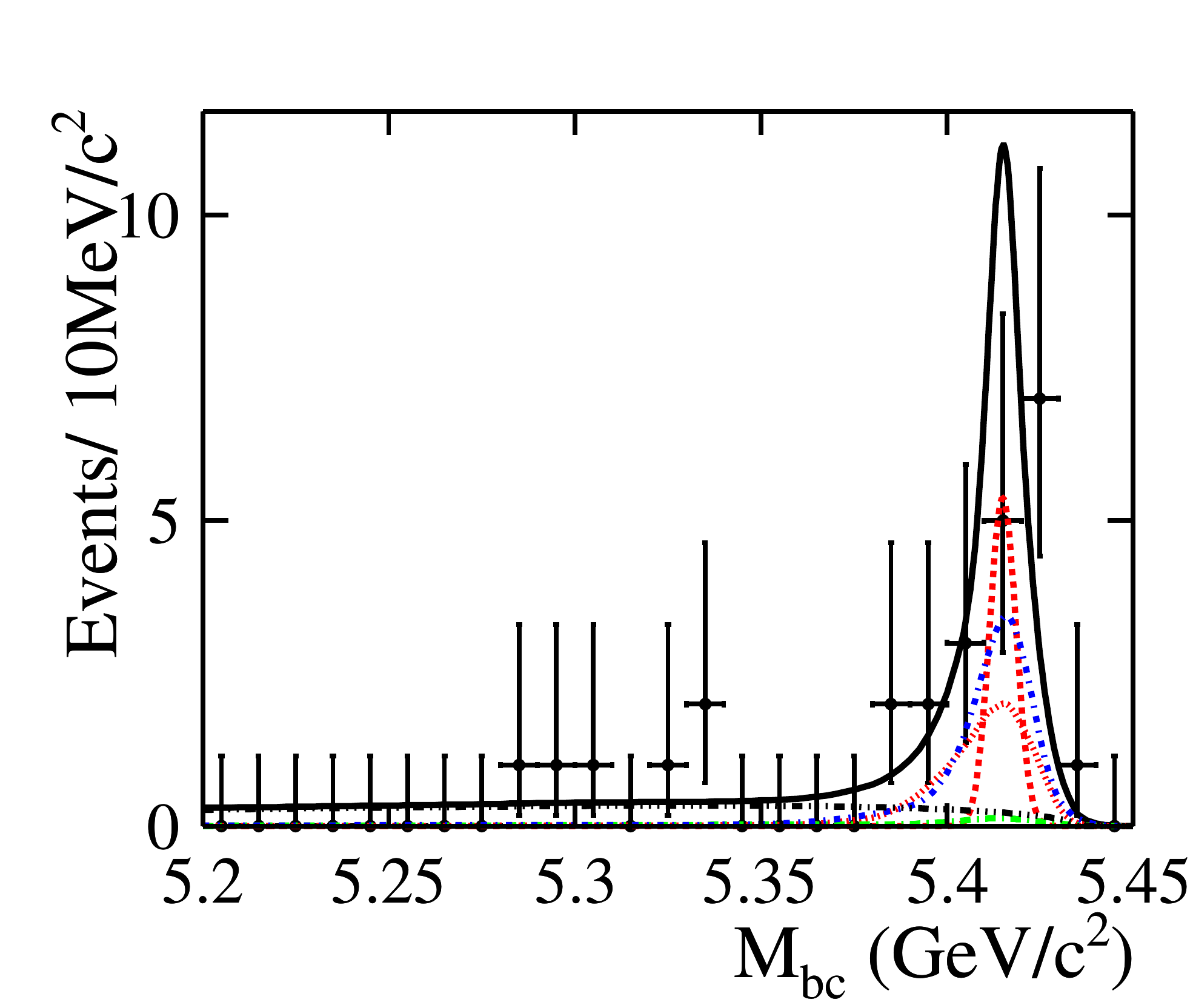}
\caption{$\Delta E$ (top) and $M_{bc}$(bottom) distributions for $D_{s}^{-}D_{s}^{+}$, $D_{s}^{-}D_{s}^{+}$ and $D_{s}^{-}D_{s}^{+}$, from left to right respectively. The red-dashed curve represents correctly reconstructed signal events, the black curve is the total fit.}
\label{fig:dsds}
\end{center}
\end{figure}

For this study \cite{SE105}, $D_{s}^{+}$ candidates are reconstructed in six modes,  $\phi\pi^{+}$, $K_{S}K^{+}$, $K^{*0}K^{+}$,  $\phi\rho^{+}$, $K^{*+}K_{S}$ and $K^{*+}K^{*0}$. $B_{s}^{0}$ candidates are reconstructed from two oppositely charged $D_{s}^{(*)}$ mesons. As the daughter photon of the $D_{s}^{*}$ has very low momentum, more than half of the events yield more than one $B_{s}^{0}$ candidate sharing the same $D_{s}$ pair. Only one candidate per event is selected  using a selection criteria based on  $M_{D_{s}}$ and $M_{D_{s}^{*}}-M_{D_{s}}$ information.  After rejecting continuum events using a Fisher discriminant based on a set of modified Fox-Wolfram moments \cite{R2, KLR}, the remaining background events are largely $B_{(s)} \to D_{s}^{(*)}X$ decays, where X is an accidental particle combination with a reconstructed mass within the $D_{s}$ mass window. The $B_{s}^{0} \to D_s^{-} D_{s}^{+} $, $D_s^{*-} D_{s}^{+} $, and $D_s^{*-} D_{s}^{*+} $ modes are fitted simultaneously; the fit projections are shown in Figure \ref{fig:dsds}.

The signal yields, branching fractions, and resulting value of $\Delta\Gamma/\Gamma_{CP}$ are  listed in Table \ref{tab:results}. Various systematic uncertainties are studied, and the resulting systematic errors are listed after the statistical errors. The second systematic error is due to uncertainty of $f_{s}$ for $B_{s}^{0}\to D_{s}^{*-}\pi^{+}$, $D_{s}^{(*)-}\rho^{+}$ modes. For $B_{s}^{0}\to D_{s}^{(*)-}D_{s}^{(*)+}$ modes, it also includes uncertainties of  $D_{s}$ branching fractions, $\sigma_{\Upsilon(5S)}$, and $f_{B_{s}^{*}\bar{B}_{s}^{*}}$. Our results are in good agreement with the theoretical predictions \cite{AD549,LN06} and existing measurements\cite{CDF-D0}.

\begin{table}[htb]
\begin{center}
\begin{tabular*}{1.0\textwidth}{@{\extracolsep{\fill}} l| c c c c | l@{\extracolsep{\fill}} } 
Mode & $N_{B_{s}^{*}\bar{B}_{s}^{*}}$ & S & $\epsilon$ & $\mathcal{B} (\%)$ & 
\textit{World Average} \\ [0.1ex]
\hline
$B_{s}^{0} \to D_{s}^{*-}\pi^{+}$ &$53.4^{+10.3}_{-9.4}$ & 7.1 & 9.13$\times 10^{-2}$  & $0.24^{+0.05} _{-0.04} \pm 0.03 \pm 0.04$  & $1^{st}$ Measurement\\  [0.1ex]
$B_{s}^{0} \to D_{s}^{-}\rho^{+}$&$92.2^{+14.2}_{-13.2}$ & 8.2 & 4.40$\times 10^{-2}$  & $0.85^{+0.13} _{-0.12} \pm 0.11 \pm 0.13$  & $1^{st}$ Measurement\\  [0.1ex]
$B_{s}^{0} \to D_{s}^{*-}\rho^{+}$&$77.8^{+14.5}_{-13.4}$ &7.4 & 2.67$\times 10^{-2}$  & $1.19^{+0.22} _{-0.20} \pm 0.17 \pm 0.18$ & $1^{st}$ Measurement\\  [0.1ex]
$f_{L}(B_{s}^{0} \to D_{s}^{*-}\rho^{+})$&  \multicolumn{3}{c}{$1.05^{+0.08}_{-0.10}{}^{+0.03}_{-0.04}$}& & $1^{st}$ Measurement \\  [0.1ex]
\hline
$B_s^0 \to D_s^{-}D_s^{+}$    &  $8.5\,^{+3.2}_{-2.6}$ & $6.2$ & 3.31$\times 10^{-4}$  & $1.03\,^{+0.39}_{-0.32}\,^{+0.15}_{-0.13}\pm0.21$  & $( 1.04 \pm 0.35 )\%$	\\[0.1ex]
$B_s^0 \to D_s^{*-}D_s^{+}$  &  $9.2\,^{+2.8}_{-2.4}$ & $6.6$ & 1.35$\times 10^{-4}$  & $2.75\,^{+0.83}_{-0.71}\pm0.40\pm0.56$ &$1^{st}$ Observation\\ [0.1ex]
$B_s^0 \to D_s^{*-}D_s^{*+}$ & $4.9\,^{+1.9}_{-1.7}$ & $3.1$ & 0.643$\times 10^{-4}$  & $3.08\,^{+1.22}_{-1.04}\,^{+0.57}_{-0.58}\pm0.63$  & $1^{st}$ Evidence\\ [0.1ex] 
$B_s^0 \to D_s^{(*)-}D_s^{(*)+}$& $22.6\,^{+4.7}_{-3.9}$ &  &  & $6.85\,^{+1.53}_{-1.30}\pm1.11\,^{+1.40}_{-1.41}$ & $(4.0 \pm 1.5 )\%$\\ [0.1ex] 
$\Delta\Gamma_{s}/\Delta\Gamma$ &  \multicolumn{3}{c}{$0.147^{+0.036}_{-0.030}{}^{+0.042}_{-0.041}$} & & 	
$0.080 \pm 0.030 $\\[0.1ex]
\hline
\end{tabular*}
\end{center}
\caption{Summary of the results. Signal yields in the $B_{s}^{*}\bar{B}_{s}^{*}$ production mode, $N_{B_{s}^{*}\bar{B}_{s}^{*}}$; significances, S (including systematics); total signal efficiencies, 
$\epsilon$ (including all sub-decay branching fractions); and branching fractions, $\mathcal{B}$. The first error is statistical, while the latter two are systematic and arise from internal and external sources. The significance $S = \sqrt{(-2\ln(L_{0}/L_{max}))}$, where $L_{0} (L_{max})$ are likelihood values when the signal yield is fixed to zero (floated).}
\label{tab:results}
\end{table}

\section{Conclusion}
We presented recent branching fraction measurements of $B_{s}^{0}$ decays obtained from 23.6 fb$^{-1}$ of $\Upsilon(5S)$ data recorded by the Belle experiment. Also, the longitudinal polarization fraction is measured for the  $B_{s}^{0}\to D_{s}^{*-}\rho^{+}$ mode and  $\Delta \Gamma_s^{CP}/\Gamma_s$ is estimated using  $D_s^{(*)-} D_{s}^{(*)+} $ modes. 


\end{document}